\documentclass[pra,twocolumn,amssymb]{revtex4}
\newcommand{\be}{\begin{equation}}
\newcommand{\ee}{\end{equation}}
\newcommand{\bea}{\begin{eqnarray}}
\newcommand{\eea}{\end{eqnarray}}
\usepackage{epsfig}
\usepackage{graphicx}

\begin{document}

\title{\textit{p}-band in a rotating optical lattice}

\author{R.~O.~Umucal{\i}lar}
\email{onur@fen.bilkent.edu.tr}
 \affiliation{ Department of
Physics, Bilkent University, 06800 Ankara, Turkey }
\author{M.~\"O.~Oktel}
\email{oktel@fen.bilkent.edu.tr} \affiliation{ Department of
Physics, Bilkent University, 06800 Ankara, Turkey }

\date{\today}

\begin{abstract}
We investigate the effects of rotation on the excited bands of a
tight-binding lattice, focusing particularly on the first excited
(\textit{p}-) band. Both the on-site energies and the hopping
between lattice sites are modified by the effective magnetic field
created by rotation, causing a non-trivial splitting and magnetic
fine structure of the \textit{p}-band. We show that Peierls
substitution can be modified to describe \textit{p}-band under
rotation, and use this method to derive an effective Hamiltonian.
We compare the spectrum of the effective Hamiltonian with a first
principles calculation of the magnetic band structure and find
excellent agreement, confirming the validity of our approach. We
also discuss the on-site interaction terms for bosons and argue
that many-particle phenomena in a rotating \textit{p}-band can be
investigated starting from this effective Hamiltonian.
\end{abstract}

\maketitle

\section{Introduction}

 Ultra-cold atom experiments display amazing versatility and
 promise to improve our understanding of many particle physics.
 There is now hope for direct experimental realization of many models
 which were constructed as effective models of condensed matter
 systems, such as the Hubbard model \cite{Esslinger1}. However, the extent of the
 ultra-cold atom experiments are not limited to previously discussed
 models, novel systems such as dipolar \cite{Pfau} and spinor gases \cite{Ketterle}
 are created as well. Through the interplay of these experiments and theories
 aimed at explaining or stimulating them, a better understanding of
 quantum many particle physics emerges.

One of the problems that has been discussed extensively in the
condensed matter literature, but never experimentally realized is
the effect of a periodic potential under high magnetic fields
\cite{Hofstadter}. When the magnetic flux through the unit cell of
the periodic potential is of the order of one flux quantum, the
energy spectrum displays complex magnetic fine structure within
the Bloch bands. Conventional condensed matter systems with
lattice constants in the order of nanometers require thousands of
tesla magnetic fields to be able to see these effects, which is
very far from being experimentally feasible.

Recently a number of authors have argued that this model can be
realized in a cold atom set-up within current experimental
capability
\cite{Wu1,Polini,Jaksch,Sorensen,Mueller,Osterloh,Ruseckas}. As
the atoms are neutral the magnetic field is expected to be created
either by rotating the optical lattice, or by optically induced
potentials. A weak rotating optical lattice has already been
realized in vortex pinning experiments \cite{Cornell}, and there
is ongoing work about creating effective magnetic fields via light
induced coupling \cite{Phillips}. Theoretical studies of these
systems promise interesting phenomena such as lattice quantum Hall
effects \cite{Palmer,Lukin,Goldman,Oktel1} or the observation of
topological conductance quantization \cite{Oktel2}.

Most of the recent theory, as well as the previous investigations
in the condensed matter literature are focused on the lowest
(\textit{s}-) band of the lattice. An important reason for this
focus is that the magnetic fine structure of this lowest band is
very well described by the Peierls substitution \cite{Obermair}.
The resulting band structure for the split \textit{s}-band is
easily obtained from a difference equation and is a self-similar
fractal, known as the Hofstadter butterfly \cite{Hofstadter}. The
original tight-binding Hamiltonian is modified only by the
addition of phases to hopping amplitudes and serves as a starting
point for the investigation of many-body physics in this system.

An exciting development in cold atom physics has been the
realization that higher bands in an optical lattice are also
experimentally accessible \cite{Esslinger2,Bloch}. The physics of
the first excited band, the \textit{p}-band, contains surprises
such as Bose condensation at non-zero momentum \cite{Liu}, or
orbitally ordered Mott insulators \cite{Girvin,Das
Sarma,Liu2,Wu2}. For a system of fermions \textit{p}-band physics
can be accessed trivially by filling the \textit{s}-band
completely; surprisingly the relaxation time for bosons in the
\textit{p}-band is long enough to allow experimental access to
pure \textit{p}-band physics.

A natural question to ask about the \textit{p}-band physics is how
the particles in the \textit{p}-band respond to the effective
magnetic field created by rotation. Experimentally, if a strong
rotating optical lattice is realized, the \textit{p}-band should
be as accessible as the \textit{s}-band. One can imagine the
already rich physics of the \textit{p}-band \cite{Das
Sarma2,Zhai,Das Sarma3,Das Sarma4,Wang,Wu3,Wu4} to be strongly
affected by the magnetic field, as both the orbital order within
each lattice site and the hopping between different lattice sites
will be modified. Beyond the single particle physics, it is not
clear how the various many particle phases, such as orbitally
ordered Mott insulators, will be affected by rotation.

The theoretical investigation of such effects requires a
consistent method of incorporating the phases generated by the
magnetic field into the lattice Hamiltonian. For the
\textit{s}-band, Peierls substitution, in which one builds an
effective Hamiltonian by replacing $\textbf{k}$ with
$(\textbf{p}-e\textbf{A}/c)/\hbar$ in the energy band function
gives a satisfactory description of the one particle physics
\cite{Hofstadter}. Starting from this effective Hamiltonian
interaction effects can be investigated. The accuracy of Peierls
substitution for the \textit{s}-band has been checked by numerical
solutions of the Schr\"{o}dinger equation \cite{Obermair}.
However, as for degenerate bands (of which the \textit{p}-band is
the simplest example) the conjecture was that ``wherever the
unperturbed Bloch bands touch or overlap, it is not possible to
obtain the magnetic sub-structure by semiclassical methods, even
approximately, by means of a universal rule for the whole
Brillouin zone" \cite{Obermair}.

In this paper, we generalize the Peierls substitution procedure to
the \textit{p}-band, and obtain an effective Hamiltonian for the
\textit{p}-band of the rotating optical lattice. We show that
after an appropriate diagonalization in $\textbf{k}$ space, which
assumes temporarily that only the on-site energies are affected by
the degeneracy lifting field, Peierls substitution is still a good
option to obtain the detailed magnetic fine structure. We check
the spectrum obtained from the effective Hamiltonian with an
accurate numerical solution of the two-dimensional Schr\"{o}dinger
equation and obtain excellent agreement. This method should in
principle be applicable to other degenerate bands and it provides
us with a means to examine inter-particle interactions.

The paper is organized as follows: In the next section, we
introduce the Hamiltonian for a rotating optical lattice, and
discuss the tight binding limit. Section III contains a discussion
of the Peierls substitution scheme, the resulting magnetic fine
structure, and its comparison with direct numerical solutions. In
Section IV, we give the effective Hamiltonian including
interactions and conclude in section V.

\section{The Model}
We start with the Hamiltonian for a particle in the rotating frame
of a two-dimensional square lattice
\begin{eqnarray} \label{Hamiltonian}
    H &=& \frac{1}{2m}\mathbf{p}_{\bot}^2 +
    \frac{1}{2}m\omega_{\bot}^2r^2 - \Omega \mathbf{\hat{z}}\cdot
    \mathbf{r}\times \nonumber
    \mathbf{p}_{\bot} \\ &+& V_0\big[\sin^2(k x)+\sin^2(k y)\big],
\end{eqnarray}
where $\mathbf{p}_{\bot} = (p_x,p_y)$ and $\mathbf{r} = (x,y)$.
$m$ is the mass of the particle, $\omega_{\bot}$ is the transverse
harmonic trapping frequency, $\Omega$ is the rotation frequency,
and $V_0$ is the depth of the optical potential created by a laser
beam with wave number $k = 2\pi/\lambda$ (for counter-propagating
laser beams lattice constant $a$ is equal to $\lambda/2$). In what
follows, we use photon recoil energy $E_{\text{R}} = \hbar^2
k^2/(2m)$ as the energy unit. This Hamiltonian can be rearranged
as

\begin{eqnarray}\label{Hamiltonian2}
    H &=& \frac{(\mathbf{p}_{\bot}-m\Omega \mathbf{\hat{z}}\times \mathbf{r})^2}{2m}
    +\nonumber
    V_0\big[\sin^2(k x)+\sin^2(k y)\big] \\ &+&
    \frac{1}{2}m(\omega_{\bot}^2-\Omega^2)r^2.
\end{eqnarray}
We neglect the last term assuming that $\Omega$ is very close to
$\omega_{\bot}$, so essentially we deal with a particle under an
effective magnetic field $B = 2mc\Omega/e$ in a lattice potential.
We assume that $V_0$ is deep enough for a tight-binding
description to apply to the system and furthermore concentrate on
the dynamics of the particles in the first excited ($p$-) band of
the lattice. Our approach is to first cast this Hamiltonian into a
second quantized form which includes the anisotropic hopping
between nearest neighbor sites, the on-site zero point energies,
and also the shift caused by rotation. Not only do we expect the
hopping between lattice sites to be affected, as it was for the
\textit{s}-band, but also the on-site energies to be modified.
However, since the hopping and on-site Hamiltonians do not
commute, a common transformation that accounts for both
modifications cannot be found.

To overcome this difficulty, we temporarily assume that the
hopping amplitudes are not affected by the effective magnetic
field and the only change is in the on-site energies. Our
expectation is that in this way we will obtain two non-degenerate
bands to which we can apply Peierls substitution. This procedure
is rather \textit{ad hoc} the validity of which is later checked
through a comparison with the first-principles results presented
previously \cite{Obermair} and reproduced here partially.

\begin{figure}
\includegraphics[scale=0.4]{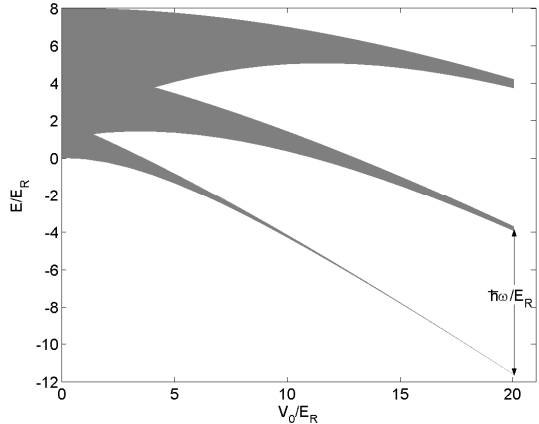}
\caption{Lowest three bands for the 2-dimensional sinusoidal
lattice potential. The energy difference between the lowest two
bands (\textit{s} and degenerate \textit{p} levels) (measured from
the band centers) is $\hbar \omega$ within the harmonic oscillator
approximation for the potential minima, $\omega$ being the
oscillator frequency. For $V_0 = 20 E_R$, $\hbar \omega = 7.7739
E_R$.} \label{fig:2_d}
\end{figure}

We proceed with considering the following \textit{p}-band
tight-binding Hamiltonian (the energy spectrum of which is
measured relative to the center of the tight-binding
\textit{s}-band) for non-interacting particles including the
on-site zero-point energies and the rotation term ($-\Omega L_z$)
\cite{Liu,Wu4}
\begin{eqnarray}\label{Hamiltonian3}
  H &=& \sum_{\textbf{R},\mu,\nu}t_{\mu \nu}(b^\dag_{\mu,\textbf{R}+a\textbf{e}_\nu} b_{\mu
  \textbf{R}}+\textrm{H.C.})+\hbar \omega \sum_{\textbf{R},\mu} b^\dag_{\mu \textbf{R}}b_{\mu
  \textbf{R}}\nonumber
   \\ &+& i\hbar\Omega \sum_{\textbf{R}}(b^\dag_{x \textbf{R}}b_{y \textbf{R}}-b^\dag_{y \textbf{R}}b_{x \textbf{R}}),
\end{eqnarray}
where the summation is over all lattice sites $\textbf{R}$ and
band indices $\mu = x,y$ (since the problem is two-dimensional,
$p_z$ orbital will not be considered). As usual, $b^\dag_{\mu
\textbf{R}}$ ($b_{\mu \textbf{R}}$) is the creation (annihilation)
operator for a particle in the $p_{\mu}$ band at lattice site
$\textbf{R}$, $\textbf{e}_\nu$ is the unit vector along the $\nu$
direction, $\omega$ is the frequency of the isotropic harmonic
oscillator potential which models the lattice potential around its
minima, and $t_{\mu \nu}$ is the anisotropic hopping amplitude.
The explicit expression for $t_{\mu \nu}$ (in the absence of
rotation) is
\begin{eqnarray}\label{t_mu_nu}
t_{\mu \nu} &=& \int
\phi^\ast_{p_{\mu}}(\textbf{r})\big[\frac{-\hbar^2\nabla^2}{2m}+V(\textbf{r})\big]\phi_{p_{\mu}}(\textbf{r}+a\textbf{e}_\nu)d\textbf{r}
\nonumber \\  &\equiv& t_\parallel \delta_{\mu \nu}-(1-\delta_{\mu
\nu})t_\bot,
\end{eqnarray}
where $V(\textbf{r})$ is the periodic lattice potential and
$\phi_{p_{\mu}}(\textbf{r})$ is the localized Wannier function
corresponding to the $p_{\mu}$ band. When we approximate the
lattice potential by a harmonic oscillator around a minimum, these
can be expressed as a product of harmonic oscillator
eigenfunctions, \textit{i.e.} $\phi_{p_{x}}(\textbf{r}) =
u_1(x)u_0(y)$ and $\phi_{p_{y}}(\textbf{r}) = u_0(x)u_1(y)$,
$u_n(x)$ being the $n^{\textrm{th}}$ harmonic oscillator
eigenfunction. $t_\parallel$ is the hopping amplitude between two
neighboring \textit{p} orbitals aligned along the orbital
orientation and $t_\bot$ is the amplitude when the orbitals are
oriented transversely with respect to the line connecting them.
Both amplitudes are defined to be positive and $t_\parallel \gg
t_\bot$ due to larger overlap. Since the lattice potential is
separable in \textit{x} and \textit{y} coordinates, $t_\parallel$
and $t_\bot$ indeed have simple expressions in reference to the
one-dimensional problem. $t_\bot$ and $t_\parallel$ are one
quarter of the widths of the lowest and next lowest bands for $V =
V_0\sin^2(k x)$, respectively. By solving the Schr\"{o}dinger
equation numerically, we find $t_\bot = 0. 0025 E_R$ and
$t_\parallel = 0.0603 E_R$ for $V_0 = 20 E_R$. The on-site
zero-point energy $\hbar\omega$ also has the simple interpretation
of being the energy difference between \textit{s} and \textit{p}
levels (bearing in mind the harmonic description, see Fig.
\ref{fig:2_d}).

\section{Peierls Substitution and Magnetic Fine Structure}

We perform a Fourier transformation on the Hamiltonian [Eq.
(\ref{Hamiltonian3})] as a preliminary for diagonalization in
momentum space. The transformed Hamiltonian is
\begin{eqnarray}\label{Hamiltonian_Fourier}
  H &=& \sum_{\textbf{k}}\big[(\epsilon_{x \textbf{k}}+\hbar\omega)b^\dag_{x \textbf{k}}b_{x \textbf{k}}+
  (\epsilon_{y \textbf{k}}+\hbar\omega)b^\dag_{y \textbf{k}}b_{y \textbf{k}}\nonumber
   \\ &+& i\hbar\Omega(b^\dag_{x \textbf{k}}b_{y \textbf{k}}-b^\dag_{y \textbf{k}}b_{x \textbf{k}})\big],
\end{eqnarray}
where $\epsilon_{\mu \textbf{k}} = 2\sum_\nu t_{\mu \nu}\cos(k_\nu
a)$. Since the Hamiltonian is bilinear in creation and
annihilation operators, it is diagonalizable by a Bogoliubov
transformation. Defining $f_{1\textbf{k}} \equiv \epsilon_{x
\textbf{k}}+\hbar\omega$ and $f_{2\textbf{k}} \equiv \epsilon_{y
\textbf{k}}+\hbar\omega$, we observe that the Hamiltonian is
diagonalized in \textbf{k} space by the following transformation:
\begin{eqnarray}\label{Bogoliubov}
\alpha_\textbf{k} = \frac{1}{\sqrt{2}}\big[
(\cos\theta_\textbf{k}+\sin\theta_\textbf{k})b_{x
\textbf{k}}+i(\cos\theta_\textbf{k}-\sin\theta_\textbf{k})b_{y
\textbf{k}}\big]
\nonumber \\
\beta_\textbf{k} = \frac{1}{\sqrt{2}}\big[
(\cos\theta_\textbf{k}-\sin\theta_\textbf{k})b_{x
\textbf{k}}-i(\cos\theta_\textbf{k}+\sin\theta_\textbf{k})b_{y
\textbf{k}}\big],
\end{eqnarray}
with
\begin{eqnarray}
\cos2\theta_\textbf{k} &=&
\frac{1}{\sqrt{1+\big(\frac{f_{1\textbf{k}}-f_{2\textbf{k}}}{2\hbar\Omega}\big)^2}},
\nonumber \\\sin2\theta_\textbf{k} &=&
\frac{f_{1\textbf{k}}-f_{2\textbf{k}}}{2\hbar\Omega}
\frac{1}{\sqrt{1+\big(\frac{f_{1\textbf{k}}-f_{2\textbf{k}}}{2\hbar\Omega}\big)^2}}.\nonumber
\end{eqnarray}
The diagonal Hamiltonian has the form
\begin{eqnarray}\label{diagonal}
H =
\sum_{\textbf{k}}\big[E_{\alpha}(\textbf{k})\alpha^\dag_\textbf{k}\alpha_\textbf{k}+E_{\beta}(\textbf{k})\beta^\dag_\textbf{k}\beta_\textbf{k}\big],
\nonumber
\end{eqnarray}
with
\begin{eqnarray}\label{dispersion}
E_{\alpha, \beta}(\textbf{k}) =
\frac{f_{1\textbf{k}}+f_{2\textbf{k}}}{2} \pm \hbar \Omega
\sqrt{1+\big(\frac{f_{1\textbf{k}}-f_{2\textbf{k}}}{2\hbar\Omega}\big)^2},
\end{eqnarray}
where upper (lower) sign refers to $\alpha$ ($\beta$). From this
point on, we can apply Peierls substitution to the dispersion
relation [Eq. (\ref{dispersion})] to obtain an operator out of it,
\textit{i.e.} we change $\textbf{k}$ to
$(\textbf{p}-e\textbf{A}/c)/\hbar$ using the Landau gauge
$\textbf{A} = Bx\hat{\textbf{y}}$. The resulting Hamiltonian is
transparent only when expressed in terms of a power series
\begin{eqnarray}\label{dispersion_series}
E_{\alpha, \beta}(\textbf{k})\!\! & = & \!\!
\frac{f_{1\textbf{k}}+f_{2\textbf{k}}}{2} \pm \hbar \Omega
\big[1+\frac{1}{2}\big(\frac{f_{1\textbf{k}}-f_{2\textbf{k}}}{2\hbar\Omega}\big)^2\\
&-&\frac{1}{8}\big(\frac{f_{1\textbf{k}}-f_{2\textbf{k}}}{2\hbar\Omega}\big)^4+...\big],\nonumber
\end{eqnarray}
with the assumption that
$|f_{1\textbf{k}}-f_{2\textbf{k}}|/2\hbar\Omega =
|\cos(k_xa)-\cos(k_ya)|(t_{\parallel}+t_{\bot})/\hbar\Omega$ is
smaller than one. If $(t_{\parallel}+t_{\bot})/\hbar\Omega$ is
much smaller than one, terms of lower order in
$(f_{1\textbf{k}}-f_{2\textbf{k}})/2\hbar\Omega$ will be more
dominant and one needs to consider only few terms for a desired
accuracy, instead of summing the whole series. Increasing accuracy
is achieved by adding higher order terms. In a typical
experimental condition, for instance, with $V_0 = 20E_R$ and
$\hbar\Omega \sim E_R$, the ratio
$(t_{\parallel}+t_{\bot})/\hbar\Omega$ is $\sim 0.063$, so a first
order approximation may be sufficient for the desired accuracy.
Here, we give the results to second order in
$(f_{1\textbf{k}}-f_{2\textbf{k}})/2\hbar\Omega$, for
completeness. The approximate energy band functions, where we
retain terms up to second order, are then
\begin{eqnarray}\label{dispersion_approximate}
E_{\alpha, \beta}(\textbf{k})\!\! & = & \!\!
\frac{f_{1\textbf{k}}+f_{2\textbf{k}}}{2} \pm \hbar \Omega
\big[1+\frac{1}{2}\big(\frac{f_{1\textbf{k}}-f_{2\textbf{k}}}{2\hbar\Omega}\big)^2\big]
\nonumber \\\!\! & = & \!\!c^{\pm}_0 +
c_1\big[\cos(k_xa)+\cos(k_ya)\big] \nonumber
\\\!\! &+&
\!\!c^{\pm}_2\big[\!\cos^2(k_xa)\!+\!\cos^2(k_ya)\!-\!2\cos(k_xa)\cos(k_ya)\!\big],\hskip
14pt
\end{eqnarray}
where $c^{\pm}_0=\hbar(\omega \pm \Omega)$, $c_1 =
t_\parallel-t_\bot$, and $c^{\pm}_2=\pm
(t_\parallel+t_\bot)^2/2\hbar\Omega$. After converting cosines
into sums of exponentials and making the Peierls substitution we
obtain discrete translation operators, which allow us to express
the eigenvalue problem as a difference equation. Since
translations along $y$ are multiplied by phases depending on $x$
in the Landau gauge, one should be careful in creating an operator
from cross terms such as $\exp(ik_xa)\exp(ik_ya)$. The correct way
of transforming should yield Hermitian operators and is obtained
by symmetric combinations such as
\begin{eqnarray}\label{symmetrization}
e^{ik_xa}e^{ik_ya} \!\!\rightarrow\!
\frac{e^{ip_xa/\hbar}e^{i(p_y-eBx)a/\hbar}+e^{i(p_y-eBx)a/\hbar}e^{ip_xa/\hbar}}{2}.\hskip
0.1 in
\end{eqnarray}
Due to the translational invariance of the problem along $y$
direction, the $y$ dependent part of the wave function is a plane wave \cite{Hofstadter}
\begin{eqnarray}\label{wavefunction}
\psi(x,y) = e^{ik_y y}g(x).
\end{eqnarray}
Making the substitutions $x = na$ and $y =la$, $n$ and $l$ being
integers, and acting the effective Hamiltonian
$E_{\alpha,\beta}[(\textbf{p}-eBx\hat{\textbf{y}}/c)/\hbar]$ on
the wave function [Eq. (\ref{wavefunction})], we get the following
difference equation
\begin{widetext}
\begin{eqnarray}\label{difference}
\frac{c^{\pm}_2}{4}\big[g(n+2)+g(n-2)\big] &+& \bigg\{
\frac{c_1}{2}-\frac{c^{\pm}_2}{2}\big[\cos(2\pi
n\phi-k_ya)+\cos(2\pi (n+1)\phi-k_ya)\big]\bigg\}g(n+1)\nonumber
\\ &+& \bigg\{\frac{c_1}{2}-\frac{c^{\pm}_2}{2}\big[\cos(2\pi
n\phi-k_ya)+\cos(2\pi (n-1)\phi-k_ya)\big]\bigg\}g(n-1)\nonumber
\\ &+& \bigg[\frac{c^{\pm}_2}{2}\cos(4\pi n\phi-2k_ya)+c_1\cos(2\pi
n\phi-k_ya)+c^{\pm}_0+c^{\pm}_2\bigg]g(n) = E g(n),
\end{eqnarray}
\end{widetext}
where $c^{\pm}_0$, $c_1$, and $c^{\pm}_2$ were introduced
following Eq. (\ref{dispersion_approximate}) and $\phi =
a^2B/(hc/e)$ is the magnetic flux quantum per unit cell. $\phi$
can be expressed in terms of the rotation frequency $\Omega$ as
$\phi = 2ma^2\Omega/h$.

\begin{figure}
\includegraphics[scale=0.4]{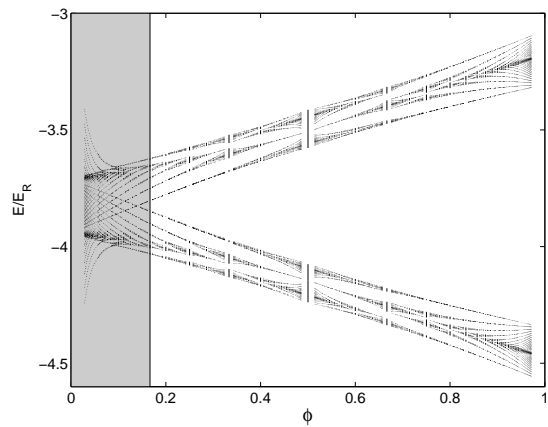}
\caption{Magnetic fine structure of the $p$-band for $V_0 = 20
E_R$. Two-fold degenerate zero-field $p$-band is split into two as
$\phi =p/q$ grows. Each split band further has $q$ sub-bands. Our
approximation fails in the shaded region, corresponding to $\phi
\lesssim 1/6$ (the nearest $\phi = 1/q$ to $1/5$, for which the
spectrum is displayed in Fig. \ref{fig:alpha1over5}), where two
bands overlap. This region can be made narrower if the lattice
depth $V_0$ is increased.} \label{fig:hofspband}
\end{figure}

When $\phi = p/q$, $p$ and $q$ being relatively prime integers,
the difference equation [Eq. (\ref{difference})] yields $q$
equations together with the Bloch condition $g(n+q)=e^{ik_x q
a}g(n)$ due to the $q$-site translational invariance in the $x$
direction. By diagonalizing the resulting $q\times q$ coefficient
matrix for several $k_x$ and $k_y$ pairs, we obtain the energy
eigenvalues which are plotted in Fig. \ref{fig:hofspband} as a
function of $\phi$. We observe that each split band further
divides into $q$ sub-bands forming a pattern which has close
resemblance to the Hofstadter butterfly. This result is in fact
anticipated since $c^{\pm}_2=\pm
(t_\parallel+t_\bot)^2/2\hbar\Omega$ is much smaller than $c_1 =
t_\parallel-t_\bot$ and if we simply neglect it as a first
approximation, the energy band function [Eq.
(\ref{dispersion_approximate})] will just be that of the
tight-binding $s$-band, except that we have $\hbar\Omega$ which
gives rise to increasing separation between the split $p$-bands
with increasing $\phi$. Our approximation becomes poorer as $\phi$
(or $\Omega$) becomes smaller since we require that
$(t_{\parallel}+t_{\bot})/\hbar\Omega$ be small. This is apparent
in Fig. \ref{fig:hofspband} in which we highlight the region where
two bands overlap. However, if we increase the lattice depth,
which decreases the hopping amplitudes, we can increase the region
of validity. Equivalently, we can say that our results should
improve as $\phi$ increases. Another improvement option would be
to consider a higher order expansion in translation operators,
which models long-range hopping with yet smaller amplitudes.

To be able to judge the accuracy of the magnetic fine structure
obtained by our method we compare it with a direct numerical
solution of the Schr\"{o}dinger equation, starting from the
Hamiltonian [Eq. (\ref{Hamiltonian2})]. One method of numerical
solution is to reduce the problem to a magnetic unit cell using
magnetic translation symmetry and solve the two dimensional
Schr\"{o}dinger equation within this unit cell using finite
difference methods. Unfortunately, the magnetic unit cell size
increases with $q$, the denominator of the flux $\phi=p/q$, and
the non-trivial boundary conditions required by magnetic
translation symmetry makes this direct solution method
computationally inefficient. Another, more efficient method, which
was first developed by Zak \cite{Zak}, and then expanded on by
Obermair \textit{et. al.} \cite{Obermair}, is to use magnetic
translation symmetry to reduce the two dimensional Schr\"{o}dinger
equation to a set of $p$ one-dimensional equations with non-local
couplings. This equation can be handled with relative ease using a
truncated basis of harmonic oscillator wave functions. Still, a
numerical calculation is efficient only for pure cases with $\phi
= 1/q$ and for small $q$ values.

\begin{figure}
\includegraphics[scale=0.4]{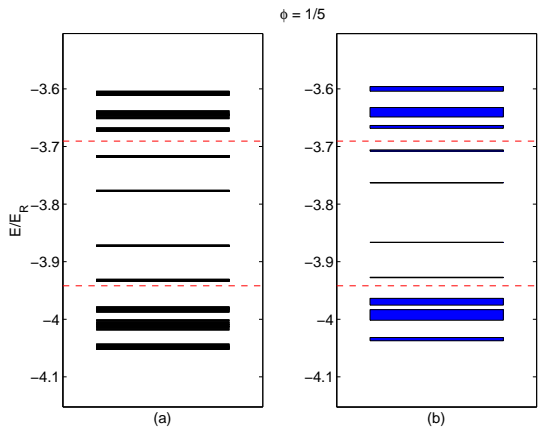}
\caption{(a) Approximate energy levels, corresponding to $\phi =
1/5$, in our effective Hamiltonian approach. (b) Band diagram
obtained through a first-principles calculation in which a
truncated basis of harmonic oscillator wave functions is used.
Dashed lines show the edges of the zero-field $p$-band.}
\label{fig:alpha1over5}
\end{figure}

\begin{figure}
\includegraphics[scale=0.4]{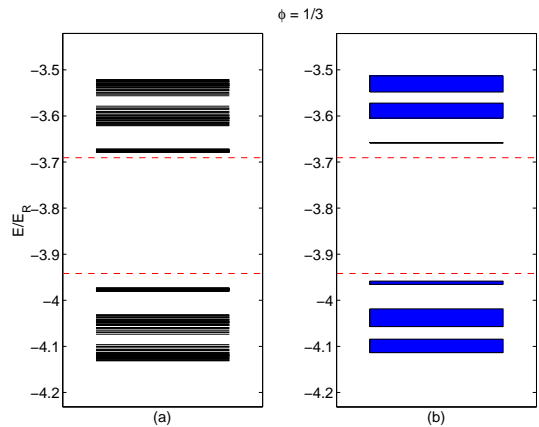}
\caption{Energy levels for $\phi = 1/3$. (a) Results of the
effective Hamiltonian approach. (b) First-principles band diagram.
Our approximation is better compared to the case of $\phi = 1/5$,
depicted in Fig. \ref{fig:alpha1over5}, in the sense that here
band gaps are also more correctly captured, apart from a slight
overall shift. Also shown, by dashed lines, are the edges of the
zero-field $p$-band.} \label{fig:alpha1over3}
\end{figure}

In Figs. \ref{fig:alpha1over5} and \ref{fig:alpha1over3}, we
compare our results with those obtained by a direct numerical
calculation along the lines of Ref. \cite{Obermair}.Calculations
with the effective Hamiltonian are much faster and the results are
as good as the direct numerical solution. For instance, in the
case of $\phi = 1/5$ (Fig. \ref{fig:alpha1over5}) the agreement is
already good, but if we increase $\phi$ to $1/3$ (Fig.
\ref{fig:alpha1over3}), apart from a slight overall shift, we see
that band gaps are also more faithfully reproduced. The
computational efficiency of the effective Hamiltonian method for
the single particle problem is striking, but its real utility is
that it can be used as a starting point to include interactions in
the system.

\section{Effective Hamiltonian}

Until now, we have essentially been dealing with the single
particle spectrum. The results we obtained can be utilized to
examine the case of many particles, if we first write the
effective Hamiltonian in real space

\begin{eqnarray}\label{Hamiltonian_real space}
\hskip -2 in H_{eff} &=& \frac{1}{4}\sum_{\langle\langle\langle
\textbf{r},\textbf{r}'\rangle\rangle\rangle}A_{\textbf{r},\textbf{r}'}(c^{+}_2\alpha^\dag_{\textbf{r}}\alpha_{\textbf{r}'}
+c^{-}_2\beta^\dag_{\textbf{r}}\beta_{\textbf{r}'})  \nonumber \\
&-&\frac{1}{4}\sum_{\langle\langle
\textbf{r},\textbf{r}'\rangle\rangle}B_{\textbf{r},\textbf{r}'}(c^{+}_2\alpha^\dag_{\textbf{r}}\alpha_{\textbf{r}'}
+c^{-}_2\beta^\dag_{\textbf{r}}\beta_{\textbf{r}'}) \nonumber \\
&+& \frac{c_1}{2}\sum_{\langle
\textbf{r},\textbf{r}'\rangle}C_{\textbf{r},\textbf{r}'}(\alpha^\dag_{\textbf{r}}\alpha_{\textbf{r}'}+\beta^\dag_{\textbf{r}}\beta_{\textbf{r}'})
\nonumber \\
&+& \sum_\textbf{r}
\big[(c^{+}_0+c^{+}_2)\alpha^\dag_{\textbf{r}}\alpha_{\textbf{r}}+(c^{-}_0+c^{-}_2)\beta^\dag_{\textbf{r}}\beta_{\textbf{r}}\big],
\end{eqnarray}
\begin{eqnarray}\label{phases}
\hskip -0.75 in A_{\textbf{r},\textbf{r}'} &=& \left\{%
\begin{array}{ll}
    e^{\pm i 4\pi n \phi}, & \hbox{$\textbf{r}$ and $\textbf{r}'$ have $x = na$;} \\
    1, & \hbox{otherwise.} \\
\end{array}%
\right. \nonumber
\end{eqnarray}
\begin{eqnarray}
B_{\textbf{r},\textbf{r}'} &=& \left\{%
\begin{array}{ll}
e^{i2\pi (\pm n+1) \phi}+e^{\pm i 2\pi n \phi}, & \hbox{$\textbf{r}$ and $\textbf{r}'$ on $y = -x$;} \\
e^{i2\pi (\pm n-1) \phi}+e^{\pm i 2\pi n \phi}, & \hbox{$\textbf{r}$ and $\textbf{r}'$ on $y = x$;} \\
&\hbox{($\textbf{r}$ or $\textbf{r}'$ has $x = na$)}.\\
\end{array}%
\right. \nonumber
\end{eqnarray}
\begin{eqnarray}
\hskip -0.68 in C_{\textbf{r},\textbf{r}'} &=& \left\{%
\begin{array}{ll}
    e^{\pm i 2\pi n \phi}, & \hbox{ $\textbf{r}$ and $\textbf{r}'$ have $x = na$;} \\
    1, & \hbox{ otherwise.} \\
\end{array}%
\right.  \nonumber
\end{eqnarray}
Here, $\langle \textbf{r},\textbf{r}'\rangle$ denotes summation
over nearest neighbors in the square lattice (with separation
$a$), $\langle\langle \textbf{r},\textbf{r}'\rangle\rangle$ over
next-nearest neighbors (with separation $\sqrt{2}a$), and
$\langle\langle\langle
\textbf{r},\textbf{r}'\rangle\rangle\rangle$ over
next-next-nearest neighbors (with separation $2a$); $\pm$ sign
refers to the hopping direction. We note that the next-nearest and
next-next-nearest coupling amplitudes turn out to be the same in
our approximation. This effective Hamiltonian represents
non-interacting particles moving in the \textit{p}-band of a
square lattice under a particular magnetic flux $\phi$. The
connection between the new and old operators is made through the
following definition
\begin{eqnarray}\label{new-old}
\alpha_\textbf{k} &\equiv& \cos\theta_\textbf{k}
b^{+}_{\textbf{k}}+\sin\theta_\textbf{k} b^{-}_{\textbf{k}}
\nonumber \\
\beta_\textbf{k} &\equiv& \cos\theta_\textbf{k}
b^{-}_{\textbf{k}}-\sin\theta_\textbf{k} b^{+}_{\textbf{k}},
\end{eqnarray}
with $b^{\pm}_{\textbf{k}} \equiv (b_{x\textbf{k}}\pm i
b_{y\textbf{k}} )/\sqrt{2}$. The operator $b^{+}_{\textbf{k}}$
($b^{-}_{\textbf{k}}$) annihilates a particle with momentum
$\hbar\textbf{k}$ whose $z$ component of angular momentum is
$-\hbar$ ($\hbar$). To first order in
$(t_{\parallel}+t_{\bot})/\hbar\Omega$, $\alpha_{\textbf{k}}$ and
$\beta_{\textbf{k}}$ are of the following form
\begin{eqnarray}\label{new-old2}
(\alpha,\beta)_\textbf{k} &=& b^{\pm}_{\textbf{k}}\pm
\frac{t_{\parallel}+t_{\bot}}{2\hbar\Omega}(\cos k_xa-\cos k_ya)
b^{\mp}_{\textbf{k}}, \nonumber
\end{eqnarray}
where the upper (lower) sign refers to $\alpha$ ($\beta$). After
expressing cosines as exponentials, we make the Peierls
substitution, \textit{i.e.} we change $\textbf{k}$ to
$\textbf{k}-eBx\hat{\textbf{y}}/\hbar c$ in the coefficients of
$b^\pm_{\textbf{k}}$ and interpret the resulting factors $\exp(\pm
i 2\pi\phi x/a)$ as momentum translation operators whose action on
a function of $\textbf{k}$ is given by $\exp(\pm i 2\pi\phi
x/a)f(\textbf{k}) = f(\textbf{k}\mp2\pi\phi \hat{\textbf{x}}/a)$.
Fourier transformation of these modified operators yields the real
space operators as
\begin{eqnarray}\label{new-old3}
\hskip -0.15 in (\alpha, \beta)_{n,l}\!\!&=&\!\!b^{\pm}_{n,l} \pm
\frac{t_{\parallel}+t_{\bot}}{4\hbar\Omega}(b^{\mp}_{n+1,l}+b^{\mp}_{n-1,l}\nonumber
\\&-& e^{i2\pi\phi n}b^{\mp}_{n,l+1}-e^{-i2\pi\phi n}b^{\mp}_{n,l-1}),
\end{eqnarray}
where the indices ($n,l$) specify the $x$ ($= na$) and $y$ ($=
la$) coordinates.

For bosons, the short-range repulsive interactions between
particles can be incorporated into our model as an on-site
interaction energy which can be written, up to terms renormalizing
the chemical potential, as \cite{Liu}
\begin{eqnarray}\label{interaction}
H_{int} &=&
\frac{U}{2}\sum_{\textbf{r}}\bigg(n^2_\textbf{r}-\frac{L^2_{z\textbf{r}}}{3\hbar^2}\bigg),\\
U &=& g\int |\phi_{p_{x,y}}(\textbf{r})|^4 d\textbf{r},\nonumber
\end{eqnarray}
where $n_{\textbf{r}} = \sum_{\mu}b^\dag_{\mu \textbf{r}} b_{\mu
\textbf{r}}$ is the boson number operator, $L_{z\textbf{r}} =
-i\hbar(b^\dag_{x \textbf{r}}b_{y \textbf{r}}-b^\dag_{y
\textbf{r}}b_{x \textbf{r}})$ is the $z$ component of the angular
momentum of a boson at site $\textbf{r}$, and $g>0$ is the
short-range repulsive interaction strength. The interaction
Hamiltonian can be written in a microscopically more revealing way
using $n^{\pm}_{\textbf{r}} = (b^{\pm}_{\textbf{r}})^\dag
b^{\pm}_{\textbf{r}}$. In this notation $n_{\textbf{r}} =
n^{+}_{\textbf{r}}+n^{-}_{\textbf{r}}$ and $L_{z\textbf{r}} =
-\hbar(n^{+}_{\textbf{r}}-n^{-}_{\textbf{r}})$. So the interaction
becomes

\begin{eqnarray}\label{interaction2}
H_{int} &=& \frac{2
U}{3}\sum_{\textbf{r}}\big[(n^{+}_{\textbf{r}})^2+(n^{-}_{\textbf{r}})^2+4
n^{+}_{\textbf{r}}n^{-}_{\textbf{r}}\big].
\end{eqnarray}
By adding $H_{int}$ [Eq. (\ref{interaction2})] to $H_{eff}$ [Eq.
(\ref{Hamiltonian_real space})], we obtain the Hamiltonian for
interacting bosons in the \textit{p}-band of a rotating optical
lattice.

\section{Conclusion}

We considered how the degenerate excited bands of a tight-binding
optical lattice are affected by the effective magnetic field
created by rotation. Specifically considering the first excited
($p$-) band of a two dimensional lattice, we pointed out that the
magnetic field causes not only the hopping between different
lattice sites to be modified, but also changes the on-site
energies. We showed that once the modification of the on-site
energies are explicitly taken into account, the Peierls
substitution scheme can be used to obtain an effective Hamiltonian
and the energy spectrum of the system.

The energy spectrum contains not only the splitting of the two
bands under the effective magnetic field, but also the fine
structure forming a pattern similar to the Hofstadter butterfly.
We compare the energies obtained from the Peierls substitution
procedure with a direct numerical solution of the Schr\"{o}dinger
equation, and observe that our procedure matches the numerical
solution to a very good accuracy.

The effective Hamiltonian is obtained by using a series expansion
in the ratio of the hopping parameter $t_\parallel$ ($\gg t_\bot$)
to $\hbar \Omega$, which is a small parameter for tight-binding
lattices except in the limit of very slow rotation. We carry out
this expansion to second order and the resulting effective
Hamiltonian contains hopping between all lattice sites that can be
connected by traversing two links [Eq. (\ref{Hamiltonian_real
space})].

While we performed a second order expansion, it is instructive to
display the effective Hamiltonian to first order in
$(t_\parallel+t_\bot)/\hbar\Omega$ in terms of the original
operators $b_{\mu\textbf{r}}$:

\begin{eqnarray}\label{Hamiltonian_approximate}
H_{eff}\!\! &=& \!\! \frac{t_\parallel+t_\bot}{2}\!\!\sum_{\langle
\textbf{r},\textbf{r}'\rangle}\big[C_{\textbf{r},\textbf{r}'}(b^+_{\textbf{r}})^\dag
b^-_{\textbf{r}'}(1-2\delta_{(\textbf{r})_x(\textbf{r}')_x})+
\textrm{H.C.}\big]\nonumber\\
&+&\!\!\frac{t_\parallel-t_\bot}{2}\!\!\sum_{\langle
\textbf{r},\textbf{r}'\rangle}C_{\textbf{r},\textbf{r}'}\big[(b^+_{\textbf{r}})^\dag
b^+_{\textbf{r}'}+(b^-_{\textbf{r}})^\dag
b^-_{\textbf{r}'}\big]\nonumber \\
&+&\!\!\sum_\textbf{r} \big[\hbar(\omega+\Omega)
n^+_{\textbf{r}}+\hbar(\omega-\Omega)
n^-_{\textbf{r}}\big] \nonumber \\
&=& \!\!\sum_{\textbf{r},\mu,\nu}t_{\mu
\nu}(b^\dag_{\mu,\textbf{r}+a\textbf{e}_\nu} b_{\mu
\textbf{r}}e^{\frac{i e}{\hbar
c}\int_\textbf{r}^{\textbf{r}+a\textbf{e}_\nu}\textbf{A}\cdot
d\textbf{r}'}+\textrm{H.C.})\nonumber
\\ &+& \!\!\hbar \omega \sum_{\textbf{r},\mu}
b^\dag_{\mu \textbf{r}}b_{\mu\textbf{r}}+i\hbar\Omega
\sum_{\textbf{r}}(b^\dag_{x \textbf{r}}b_{y \textbf{r}}-b^\dag_{y
\textbf{r}}b_{x \textbf{r}}),\\
\hskip -0.68 in C_{\textbf{r},\textbf{r}'} &=& \left\{%
\begin{array}{ll}
    e^{\pm i 2\pi n \phi}, & \hbox{ $(\textbf{r})_x = (\textbf{r}')_x = na$;} \\
    1, & \hbox{ $(\textbf{r})_x \neq (\textbf{r}')_x$.} \\
\end{array}%
\right.  \nonumber
\end{eqnarray}
This Hamiltonian incorporates the first non-vanishing effects of
rotation and can be used as an effective Hamiltonian if
$(t_\parallel+t_\bot)/\hbar\Omega$ is not large. Indeed a recent
preprint which appeared while this paper was in preparation uses
this form as a starting point \cite{Wu4}. However, to investigate
corrections for slower rotation one has to go to higher orders as
in Eq. (\ref{Hamiltonian_real space}). In Eq.
(\ref{Hamiltonian_approximate}), we display the vector potential
\textbf{A} explicitly to express the gauge invariance of the
effective Hamiltonian. Our numerical work was carried out using a
higher order approximation [Eq. (\ref{Hamiltonian_real space})]
which is also gauge invariant.

In conclusion, we showed how Peierls substitution can be used for
degenerate bands and checked its accuracy with direct numerical
solutions. By investigating how operator transformations are
modified through Peierls substitution [Eq. (\ref{new-old3})] we
derived a first order effective Hamiltonian in real space [Eq.
(\ref{Hamiltonian_approximate})].

Going to the next order, we obtain a more accurate, but more
complicated effective Hamiltonian, which displays how higher order
hopping is modified by the effective magnetic field. Finally, we
also give the expression for on-site interaction for bosons in
terms of the angular momentum `up' and `down' operators. We hope
that our results stimulate further theoretical and experimental
investigations of the \textit{p}-band physics under an effective
magnetic field.

\acknowledgements

R. O. U. is supported by T\"{U}B\.{I}TAK. M. \"{O}. O. is
supported by T\"{U}B\.{I}TAK-KAR\.{I}YER Grant No. 104T165 and a
T\"{U}BA-GEB\.{I}P grant.


\begin{thebibliography}{99}
\bibitem{Esslinger1}  R. J\"{o}rdens, N. Strohmaier, K. G\"{u}nter, H. Moritz, and T.
Esslinger, \textit{A Mott insulator of fermionic atoms in an
optical lattice}, arXiv:0804.4009.
\bibitem{Pfau} J. Stuhler, A. Griesmaier, T. Koch, M. Fattori, T.
Pfau, S. Giovanazzi, P. Pedri, and L. Santos, Phys. Rev. Lett.
\textbf{95}, 150406 (2005).
\bibitem{Ketterle} J. Stenger, S. Inouye, D.M. Stamper-Kurn, H.-J.
Miesner, A.P. Chikkatur, and W. Ketterle, Nature \textbf{396}, 345
(1998).
\bibitem{Hofstadter} D. R. Hofstadter, Phys. Rev. B \textbf{14}, 2239 (1976).
\bibitem{Wu1} C. Wu, H. Chen, J. Hu, and S.-C. Zhang, Phys. Rev. A \textbf{69},
043609 (2004).
\bibitem{Polini} M. Polini, R. Fazio, M. P. Tosi, J. Sinova, and A. H. Mac-
Donald, Laser Phys. \textbf{14}, 603 (2004).
\bibitem{Jaksch} D. Jaksch and P. Zoller, New J. Phys. \textbf{5}, 56 (2003).
\bibitem{Sorensen} A. S. S\o rensen, E. Demler, and M. D. Lukin, Phys. Rev. Lett.
\textbf{94}, 086803 (2005).
\bibitem{Mueller} E. J. Mueller, Phys. Rev. A \textbf{70}, 041603(R) (2004).
\bibitem{Osterloh} K. Osterloh, M. Baig, L. Santos, P. Zoller, and M. Lewenstein,
Phys. Rev. Lett. \textbf{95}, 010403 (2005).
\bibitem{Ruseckas} J. Ruseckas, G. Juzeliunas, P. Ohberg, and M. Fleischhauer,
Phys. Rev. Lett. \textbf{95}, 010404 (2005).
\bibitem{Cornell} S. Tung, V. Schweikhard, and E. A. Cornell, Phys. Rev. Lett.
\textbf{97}, 240402 (2006).
\bibitem{Phillips} First results were reported at the American Physical Society March Meeting
Y.-J. Lin, W. D. Phillips, J. V. Porto, and I. Spielman, Bull. Am.
Phys. Soc. 53, A14.00001 (2008).
\bibitem{Palmer} R. N. Palmer and D. Jaksch, Phys. Rev. Lett. \textbf{96}, 180407
(2006).
\bibitem{Lukin} M. Hafezi, A. S. S\o rensen, E. Demler, and M. D.
Lukin, Phys. Rev. A \textbf{76}, 023613 (2007).
\bibitem{Goldman} N. Goldman and P. Gaspard, Europhys. Lett. \textbf{78}, 60001 (2007).
\bibitem{Oktel1} R. O. Umucal{\i}lar and M. \"{O}. Oktel, Phys. Rev. A
\textbf{76}, 055601 (2007).
\bibitem{Oktel2} R. O. Umucal{\i}lar, H. Zhai, and M. \"{O}.
Oktel, Phys. Rev. Lett. 100, 070402 (2008).
\bibitem{Obermair} G. M. Obermair and H.-J. Schellnhuber, Phys. Rev. B \textbf{23}, 5185
(1981); H.-J. Schellnhuber, G. M. Obermair, and A. Rauh,
\emph{ibid.} \textbf{23}, 5191 (1981).
\bibitem{Esslinger2} M. K\"{o}hl, H. Moritz, T. St\"{o}ferle, K. G\"{u}nter, and T. Esslinger, Phys. Rev. Lett. \textbf{94}, 080403
(2005).
\bibitem{Bloch} T. M\"{u}ller, S. F\"{o}lling, A. Widera, and I.  Bloch, Phys. Rev. Lett. 99, 200405
(2007).
\bibitem{Liu} W. V. Liu and C. Wu, Phys. Rev. A \textbf{74}, 013607 (2006).
\bibitem{Girvin} A. Isacsson and S. M. Girvin, Phys. Rev. A \textbf{72},
053604 (2005).
\bibitem{Das Sarma} C. Wu, W. V. Liu, J. Moore, and S. Das Sarma, Phys. Rev. Lett. \textbf{97}, 190406
(2006).
\bibitem{Liu2} E. Zhao and W. V. Liu, Phys. Rev. Lett. \textbf{100}, 160403
(2008).
\bibitem{Wu2} C. Wu, Phys. Rev. Lett. \textbf{100}, 200406 (2008).
\bibitem{Das Sarma2} C. Wu, D. Bergman, L. Balents, and S. Das
Sarma, Phys. Rev. Lett. 99, 070401 (2007).
\bibitem{Zhai} K. Wu and H. Zhai, Phys. Rev. B \textbf{77}, 174431 (2008).
\bibitem{Das Sarma3} C. Wu and S. Das Sarma, Phys. Rev. B \textbf{77}, 235107 (2008).
\bibitem{Das Sarma4} V. M. Stojanovic, C. Wu , W. V. Liu, and S. Das
Sarma, \textit{Incommensurate superfluidity of bosons in a
double-well optical lattice}, arxiv:0804.3977.
\bibitem{Wang} L. Wang, X. Dai, S. Chen, and X. C. Xie, \textit{Magnetism of Cold Fermionic Atoms on p-Band
 of an Optical Lattice}, arXiv:0805.2719.
\bibitem{Wu3} S. Zhang and C. Wu, \textit{Proposed realization of
itinerant ferromagnetism in optical lattices}, arXiv:0805.3031.
\bibitem{Wu4} C. Wu, \textit{Orbital analogue of quantum anomalous Hall effect in p-band
systems}, arXiv:0805.3525.
\bibitem{Zak} J. Zak, Phys. Rev. \textbf{136}, A1647 (1964).
\end{thebibliography}
\end{document}